\documentstyle[12pt,epsf]{article}
\input epsf
\def\be{\begin{equation}}
\def\ee{\end{equation}}
\def\bea{\begin{eqnarray}}
\def\eea{\end{eqnarray}}
\begin{document}

\begin{flushright}
PU-RCG/98-1\\
{\sf gr-qc/9803021}\\
\today
\end{flushright}

\begin{center}
\Large
{\bf Phase-plane analysis of Friedmann-Robertson-Walker\\
cosmologies in Brans--Dicke gravity}

\vspace{.4in}
 
\normalsize
 
\large{Damien J.~Holden and David Wands}
 
\normalsize
\vspace{.3in}
 
{\em School of Computer Science and Mathematics, University of Portsmouth, \\ 
Mercantile House, Hampshire Terrace, Portsmouth, PO1 2EG, U. K.}
\end{center}
 
\vspace{.4in}

\begin{abstract} 
We present an autonomous phase-plane describing the
evolution of Friedmann-Robertson-Walker models containing a perfect
fluid (with barotropic index $\gamma$) in Brans-Dicke gravity (with
Brans-Dicke parameter $\omega$). We find self-similar fixed points
corresponding to Nariai's power-law solutions for spatially flat models
and curvature-scaling solutions for curved models. At infinite values
of the phase-plane variables we recover O'Hanlon and Tupper's vacuum
solutions for spatially flat models and the Milne universe for negative
spatial curvature.  We find conditions for the existence and stability
of these critical points and describe the qualitative evolution in all
regions of the $(\omega,\gamma)$ parameter space for $0\leq\gamma\leq2$
and $\omega>-3/2$.  We show that the condition for inflation in
Brans-Dicke gravity is always stronger than the general relativistic
condition, $\gamma<2/3$. 
\end{abstract}

\section{Introduction}
\setcounter{equation}{0}

One of the simplest extensions to the Einstein-Hilbert action of
general relativity is the introduction of a scalar field non-minimally
coupled to the metric tensor~\cite{Jordan}. The simplest scalar-tensor
theory is that proposed by Brans and Dicke~\cite{BD61} where the
gravity theory contains only one dimensionless parameter, $\omega$,
and the effective gravitational constant is inversely proportional to
the scalar field, $\Phi$. This theory yields the correct Newtonian
weak-field limit, but solar system measurements of post-Newtonian
corrections require $\omega>500$~\cite{Will93}.  In the limit
$\omega\to\infty$ the field $\Phi$ becomes fixed and we recover
Einstein gravity.  This has led to more general scalar-tensor gravity
theories~\cite{Bergmann,Wagoner,Nordtvedt} being considered with a
self-interaction potential $V(\Phi)$~\cite{BM90,Gregory} or a variable
$\omega(\Phi)$~\cite{DamourNordt} in order to fix $\Phi$ by the
present day.

The fact that such scalar-tensor gravity theories are a generic
prediction of low-energy effective supergravity theories from string
theory~\cite{string} or other higher-dimensional gravity
theories~\cite{KK} has led to considerable interest in cosmological
models of the very early universe derived from Brans-Dicke
gravity. The time-dependent gravitational constant introduces a new
degree of freedom into cosmological models which has led different
authors to propose extended inflation~\cite{extinf}, pre big
bang~\cite{pbb} and gravity-driven inflation models~\cite{mad} of the
early universe, as well as modifications to conventional
inflation~\cite{inf} or nucleosynthesis models~\cite{nuc}.

In this paper we will investigate the qualitative evolution of
Friedmann-Robertson-Walker (FRW) models containing a barotropic
perfect fluid in Brans-Dicke gravity.  In Section~\ref{sectflat} we
review the asymptotic behaviour for spatially flat FRW models and the
general solution which can be given in parametric form for barotropic
fluids~\cite{GFR73,Morganstern72}. Analytical solutions for spatially
curved models are only possible in a limited number of special cases
such as for radiation~\cite{M71a,LP,Barrow93,mad} or stiff
fluid~\cite{LP,MW95a} and other methods are required to study the
general behaviour.  In Section~\ref{sectphaseplane} the field
equations describing the cosmological evolution of the Brans-Dicke
field and scale factor are reduced to two coupled autonomous
first-order equations.  A phase-plane analysis~\cite{WE97} is then
used to solve qualitatively for the cosmological
evolution~\cite{WANDS93,Kolitch}. We present for the first time a
compactified phase-plane which represents both late and early time
behaviour for flat $(k=0)$ and curved $(k\ne0)$ FRW models. In
section~\ref{sectqual} the cosmological evolution is described
qualitatively for all regions of parameter space $(\gamma,\omega)$,
where the Brans-Dicke parameter $\omega>-3/2$ and the barotropic index
of the fluid $0\leq\gamma\leq2$. In Section~6 we consider whether it
is possible to construct cosmological solutions with infinite proper
lifetimes in some parameter regimes. Our results are summarised in
Section~7.

\section{Equations of motion}
\setcounter{equation}{0}

We will consider here the simplest scalar-tensor theory as
originally proposed by Brans and Dicke, where the gravity theory is
characterised by a single dimensionless parameter $\omega$, representing
the strength of coupling between the scalar field and the metric. The
equations of motion are derived by extremizing the action
\be
\label{BDEA}
S = \frac{1}{16\pi} \int d^4x \, \sqrt{-g} \, \left[
 \Phi R - \frac{\omega}{\Phi} g^{\mu\nu} \Phi_{,\mu} \Phi_{,\nu}
 + 16\pi {\cal L}_{\rm matter} \right] \ .
\ee
Homogeneous and isotropic cosmologies can be
described by the Friedmann-Robertson-Walker (FRW) metric, 
\be
ds^2 = -dt^2 + a(t)^2 \left[ \frac{dr^2}{1-kr^2}
 + r^2 (d\theta^2 + \sin^2 \theta d\phi^2) \right] \ ,
\ee
The evolution equations for the scale factor, $a(t)$, and the Brans--Dicke
field, $\Phi(t)$, can then be written as
\bea
\label{STAEQN}
{\ddot{a} \over a} & = &
 - {8\pi \over 3\Phi} \left( {3\omega P + (\omega+3)\rho \over 2\omega
 +3} \right)
 + {\dot{a}\over a} {\dot\Phi\over\Phi} 
 - {\omega\over3} \left( {\dot\Phi\over\Phi} \right)^2 \ ,\\ 
\label{STPHIEQN}
 \ddot{\Phi} &=&
 - 3\frac{\dot{a}}{a}\dot{\Phi}
 + {8\pi ( \rho - 3P ) \over 2\omega+3} \ ,
\eea
where a dot denotes differentiation with respect to cosmic time, $t$.
Matter with density $\rho$ and pressure $P$ obeys the continuity equation
\be
\label{STCONTINUITY}
\dot{\rho} = -3 \frac{\dot{a}}{a} ( \rho + P ) \ .
\ee
These evolution equations possess a first-integral which
corresponds to the generalised Friedmann constraint for the Hubble
expansion, $H\equiv\dot{a}/a$,
\be
\label{STCONSTRAINT}
H^2 = \frac{8\pi\rho}{3\Phi}
 - H \frac{\dot{\Phi}}{\Phi}
 + \frac{\omega}{6} \left( \frac{\dot{\Phi}}{\Phi} \right)^2
 - \frac{k}{a^2} \ .
\ee

Considerable insight can be gained into the dynamical evolution of
these models by working with the conformally rescaled Einstein
metric~\cite{Dicke62,conf}
\be
\tilde{g}_{\mu\nu} = (G_N\Phi) g_{\mu\nu} \ ,
\ee
where $G_N$ is Newton's constant which remains fixed in the Einstein
frame. The rescaled scale factor becomes $\tilde{a}=(G_N\Phi)^{1/2}a$
and the cosmic time
\be
\label{Etime}
\tilde{t} = \int (G_N\Phi)^{1/2} dt \ . 
\ee
The constraint equation~(\ref{STCONSTRAINT}) then can be written as
the familiar Friedmann constraint
\be
\label{ECON}
\tilde{H}^2 = {8\pi G_N \over 3} \left( \tilde\rho + \hat\rho \right)
- {k \over \tilde{a}^2} \ ,
\ee
where $\tilde{H}$ is the Hubble expansion in the Einstein frame,
$\tilde\rho=\rho/(G_N\Phi)^2$, and
\be
\hat\rho = {2\omega+3\over12} \left( {1\over\Phi}{d\Phi\over
d\tilde{t}} \right)^2 \ ,
\ee
is the effective energy density of the Brans-Dicke field in the
Einstein frame. This has an effective pressure $\hat{P}=\hat\rho$, but
is non-minimally coupled to the barotropic fluid. Note however that
from the Friedmann constraint equation~(\ref{ECON}) we can immediately
deduce that only closed models ($k>0$) have a turning point
$\tilde{H}=0$ in the Einstein frame for $2\omega+3>0$. Theories with
$2\omega+3<0$ have a negative energy density in the Einstein frame,
implying that the Minkowski vacuum spacetime is unstable, and
so we shall not consider such models here.

An equation of state for matter, $P(\rho)$, is needed in order to
solve these equations of motion.  We restrict our
analysis to barotropic fluids where the pressure is simply
proportional to the density, $P=(\gamma -1)\rho$, which includes the
case of pressureless dust ($\gamma=1$), radiation ($\gamma=4/3$) or a
false vacuum energy density ($\gamma=0$).  The continuity equation
(\ref{STCONTINUITY}) can then be integrated to give the matter density
as a function of the cosmological scale factor $a(t)$
\be
\rho = \frac{\Gamma}{a^{3\gamma}} \\
\ee
and this allows us to eliminate both $\rho$ and $P$ from the
remaining equations of motion.

General analytic solutions for scalar--tensor cosmologies exist for
only a limited number of special cases. For example we can solve
analytically for a general FRW models in vacuum
($\rho=0$)~\cite{OHT72,Barrow93} or when the matter is radiation
($P=\rho/3$)~\cite{M71a,LP,Barrow93,mad} or a stiff fluid
($P=\rho$)~\cite{LP,MW95a} when the system reduces to perfect fluids in
the Einstein frame. In Brans--Dicke gravity the general analytic
solution for other cases can only be given in parametric form for
spatially flat models ($k=0$) and we shall review these results
briefly before going on to study the evolution of the more general FRW
models.

\section{Spatially-flat solutions}
\label{sectflat}
\setcounter{equation}{0}

Brans and Dicke's original motivation was to produce a model
where the effective gravitational constant decreases ($\Phi$ increases)
as the density of the universe decreases.
In their original paper~\cite{BD61}, Brans and Dicke gave a power-law
solution for the evolution of a spatially flat FRW model filled with
pressureless dust
\be
\label{powerlaw}
a \propto |t-t_*|^p \ , \qquad \Phi \propto |t-t_*|^q \ ,
\ee
where $p=(2\omega+2)/(3\omega+4)$, and $q=2/(3\omega+4)$.
This was generalised to arbitrary barotropic equation of state by
Nariai who found~\cite{Nariai68}
\bea
\label{NARIAIp}
p & = & \frac{2\omega (2-\gamma) +2}{3\omega\gamma (2-\gamma) +4} \\
\label{NARIAIq}
q & = & \frac{2 (4-3\gamma)}{3\omega\gamma (2-\gamma)+4} \ .
\eea
This yields the general relativistic solution, $a\propto
t^{2/3\gamma}$ with $\Phi$ constant, in the limit $\omega\to\infty$.

However Nariai's power-law solution is in fact only a particular
solution. The general solution can be given analytically in terms of a
rescaled time coordinate
\be
\label{GFRT}
dT \equiv \frac{dt}{a^{3(\gamma-1)}}
\ee
Then the scale factor and Brans--Dicke field evolve
as~\cite{GFR73,Morganstern72} 
\bea
a & = & a_0 (T-T_-)^{n_\mp} (T-T_+)^{n_\pm} \ ,\\
\Phi & = & \Phi_0 (T-T_-)^{m_\mp} (T-T_+)^{m_\pm} \ ,
\eea
where
\bea
n_{\pm} & = &
 \frac{\omega}
      {3 \left( 1+\omega(2-\gamma) \mp \sqrt{1+2\omega/3} \right)}  \\
m_{\pm} & = &
 \frac{1 \pm \sqrt{1+2\omega/3} }
      {1+\omega(2-\gamma) \mp \sqrt{1+2\omega/3} }
\eea
The solution where $T_+=T_-$ reduces to Nariai's power--law
solution given by Eqs.~(\ref{powerlaw}--\ref{NARIAIq}), 
and this is clearly the attractor solution when $T\to\infty$. 
However, substituting a solution of the form given in
Eq.~(\ref{powerlaw}) into the definition of the time coordinate $T$ in
Eq.~(\ref{GFRT}) we can see that $t\to\infty$ at a {\em finite} value of $T$
if $3(\gamma-1)p>1$. Thus Nariai's solutions are {\em not} necessarily
the attractor solutions at late cosmic times.
This case needs to be studied more carefully to determine the
attractor behaviour which we shall do in the phase-plane analysis that
follows.

As $T$ approaches $T_+$ or $T_-$ the general evolution (for $T_-\neq
T_+$) approaches the vacuum solutions ($\rho=0$) first presented by
O'Hanlon and Tupper~\cite{OHT72}, which are also power-law solutions
of the form given in Eq.~(\ref{powerlaw}) with
\bea
\label{OHpq}
p_\pm = {n_\pm \over 1+3(\gamma-1)n_\pm}
 & = & \frac{\omega+1 \pm \sqrt{1+2\omega/3}}{4+3\omega} \ , \nonumber \\
q_\pm = {m_\pm \over 1+3(\gamma-1)n_\pm}
 & = & \frac{-2}{1 \pm 3 \sqrt{1+2\omega/3}} \ .
\eea
Nariai's solution is sometimes described as the
matter-dominated ($T\to\infty$) solution as opposed to the
vacuum-dominated ($T\to T_\pm$) solution.
The vacuum-dominated evolution has two solutions which have been termed {\em
fast} or {\em slow} \cite{GFR73} depending on the early time behaviour
of the scalar field $\Phi$. The {\em slow} solution corresponds to
$(p_-,q_-)$ defined above where at early times the scalar field $\Phi$
is increasing (and $G\to0$) as cosmic time increases for
$\omega>-4/3$. The {\em fast} solution is defined by $(p_+,q_+)$, with
$\Phi$ decreasing (and $G\to\infty$) as $t$ increases for
$\omega>-4/3$.

The Brans-Dicke effective action Eq.~(\ref{BDEA}) reduces to the
truncated string effective action in the absence of matter ($\rho=0$)
when the coupling constant $\omega=-1$~\cite{string}. 
For comparison with the string literature we will define,
\be
\phi \equiv - \ln \Phi \ , \qquad \alpha= \ln a \ .
\ee
It can be shown that the spatially flat FRW solutions are related by a
``scale-factor duality''~\cite{sfd}
\be
\label{SFD}
\alpha \rightarrow -\alpha \ , 
\qquad 
\phi\rightarrow \phi - 6\alpha \ .
\ee
This scale--factor duality relates expanding cosmologies
to contracting ones. Lidsey~\cite{jim} has shown that this duality can
be extended to all values of $\omega$ for vacuum solutions in 
Brans--Dicke gravity,
\bea
\label{dualtr}
\alpha & \rightarrow & \left(\frac{2+3\omega}{4+3\omega}\right)\alpha
-2\left(\frac{1+\omega}{4+3\omega}\right) \phi \nonumber \\ 
\phi & \rightarrow & -\left(\frac{6}{4+3\omega}\right)\alpha -
\left(\frac{2+3\omega}{4+3\omega}\right) \phi \ ,
\eea
which reduces to Eq.~(\ref{SFD}) when $\omega=-1$.  Thus the {\em fast}
and {\em slow} solutions defined above are interchanged under the
above duality transform\footnote{The {\em fast}, $(p_+,q_+)$, and {\em
slow}, $(p_-,q_-)$, solutions should not to be confused with the $(+)$
and $(-)$ branches defined in the context of low-energy string
cosmology by Brustein and Veneziano~\cite{BV}.}
\be
\label{dualpq}
(p_\pm,q_\pm) \rightarrow (p_\mp,q_\mp)
\ee

\section{Autonomous phase-plane}
\label{sectphaseplane}
\setcounter{equation}{0}

In the absence of analytic solutions for general FRW models we seek a
qualitative description of the evolution of the cosmological scale
factor and scalar field with respect to
time~\cite{WE97,WANDS93,Kolitch}. We shall now show that we can reduce the
equations of motion~(\ref{STAEQN}) and~(\ref{STPHIEQN}) to a
two-dimensional autonomous phase-plane whose qualitative evolution can
be described for any values of the Brans-Dicke parameter $\omega$ or
barotropic index $\gamma$.

The trace of the energy-momentum tensor for matter acts like a
potential gradient term driving the scalar field $\Phi$ in
Eq.~(\ref{STPHIEQN}) and so, in analogy with the phase-plane analysis
of self-interacting scalar fields in general
relativity~\cite{Halliwell87,BB88,MFB92}, we can absorb the matter
terms on the right-hand side of both the field equations (\ref{STAEQN})
and (\ref{STPHIEQN}) by introducing a new time coordinate, $\tau$,
such that
\be
\label{DTAU}
d\tau \equiv \sqrt{ \frac{8\pi\rho}{(3+2\omega)\Phi} }
\, dt
\ee

We introduce two new variables~\cite{WANDS93}
\bea
\label{DEFX}
x &\equiv& \frac{1}{2} \frac{\Phi'}{\Phi} \ ,\\
\label{DEFY}
y &\equiv& \frac{a'}{a} \ .
\eea
Substitution of these variables in the equations of motion
(\ref{STPHIEQN}) and (\ref{STAEQN}) yields
\bea
\label{XPRIME}
x' & = & -x^2
 - \frac{3(2-\gamma)}{2} xy + \frac{4 - 3\gamma}{2} \ , \\
\label{YPRIME}
y' & = & {-2(1-3\alpha^2)\over3\alpha^2} x^2 + 3xy + {3\gamma-2\over2}y^2 -
 {3(4-3\gamma)\alpha^2+3\gamma-2 \over 6\alpha^2} \ ,
\eea
where for convenience we have defined
\be
\alpha^2 \equiv \frac{1}{2\omega+3} \ .
\ee
We will consider only models for which $\omega>-3/2$ and thus
$\alpha^2>0$.

The constraint equation (\ref{STCONSTRAINT}) becomes
\be
\label{EXPLICITK}
(x+y)^2 = \frac{1}{3\alpha^2} \left(1+x^2\right)
 - \frac{3\Phi k}{8\pi\rho\alpha^2 a^2}
\ee
By comparison with the Einstein frame constraint equation~(\ref{ECON})
we see that $(x+y)$ describes the expansion in the Einstein frame, and
$x^2$ the effective energy density of the Brans-Dicke field relative
to the barotropic fluid.  
The spatially flat ($k=0$) models lie on a separatrix in the phase--plane:
\be
\label{k0}
x+y = \pm \, {\sqrt{1+x^2} \over \sqrt{3}\alpha} \ .
\ee
Note that in order for $k=0$ models to have a turning point in the
scale factor ($y=0$) we require $3\alpha^2>1$, i.e., $\omega<0$.

The equations~(\ref{XPRIME}) and~(\ref{YPRIME}) are two first-order
differential equations forming a autonomous dynamical system in $x$
and $y$.  The simultaneous solution of such equations defines a set of
trajectories in the $(x,y)$ plane. The graphical representation of
these equations form a system of `integral curves' where
\be
\label{DYDX}
\frac {dy}{dx} = \frac {Q(x,y)}{P(x,y)}
\ee
where the functions $P(x,y)$ and $Q(x,y)$ are defined by the right hand
sides of (\ref{XPRIME}) and (\ref {YPRIME}) respectively.
This allows the qualitative evolution of the variables $x$
and $y$ to be determined.

\subsection{Critical points in the finite plane}

The key aspect of a phase plane analysis is to determine the critical
points $(x_i,y_i)$ for the system. These are fixed points where the time
derivatives of the variables vanish,
\be
x' = 0 \ , \qquad y' = 0 \ . 
\ee
For an integral curve approaching such a point, the critical point
represents the asymptotic behaviour either at the start or the end of
the evolution.  

Firstly we seek critical points at finite values $(x_i,y_i)$ in the
phase plane. Requiring $x'=0$ implies
\be
\label{XPRIMEZERO}
x_i^2 \left( 1 + \frac{3(2-\gamma)}{2}\frac{y_i}{x_i} \right)
 = \frac{4-3\gamma}{2} \ .
\ee
Requiring in addition that $y'=0$ generates two possible solutions:
\bea
\label{SOLA}
& (a) & \hspace{0.2in} 3\alpha^2 ( x_a + y_a )^2 = ( 1 + x_a^2 ) \\
{\rm or} \hspace{0.2in}
\label{SOLB}
& (b) & \hspace{0.2in} (2-3\gamma) y_b = 2 x_b
\eea
Simultaneous solution of Eq.~(\ref{XPRIMEZERO}) with either
Eq.~(\ref{SOLA}) or~(\ref{SOLB}) gives expressions for the
critical points:
\bea
\label{XAYA}
(x_{a\pm},y_{a\pm}) & = &
 \pm \left( \frac {\alpha(4-3\gamma)}
{\sqrt{3(2-\gamma)^2 - \alpha^2(4-3\gamma)^2 }}
 , \frac {(2-\gamma) - \alpha^2 (4-3\gamma)}{ \alpha \sqrt{
3(2-\gamma)^2 - \alpha^2(4-3\gamma)^2 } } \right) \\
\label{XBYB}
(x_{b\pm},y_{b\pm}) & = & \pm \left( \frac {\sqrt{2-3\gamma}}{2} , \frac
{1}{\sqrt{2-3\gamma}} \right)
\eea

Note that from the definitions of the variables $x$ and $y$ in
Eqs.~(\ref{DEFX}) and~(\ref{DEFY}) the phase plane is symmetric under
the $180^\circ$ rotation $x\to-x$ and $y\to-y$ plus time reversal
$\tau\to-\tau$. Thus for every critical point $x_{i+},y_{i+}$ we have
a critical point $x_{i-},y_{i-}$. In the remainder of this section we
shall refer only to the critical points with $x_{i+}+y_{i+}>0$ which
correspond to expanding solutions in the Einstein frame.  It should be
taken as read that there will be a point with $x_{i-}=-x_{i+}$ and
$y_{i-}=-y_{i+}$ which corresponds to the time reversed solutions.

Also from the definitions of $x$ and $y$ we see that the fixed points
$(x_i,y_i)$ correspond to solutions for $\Phi\propto e^{2x_i\tau}$ and
$a\propto e^{y_i\tau}$ with respect to the time coordinate $\tau$. The
evolution at the fixed points is self-similar as the nature of the
solutions is unaffected by a rescaling $\tau\to\tau+\Delta\tau$.
Integration of Eq.~(\ref{DTAU}) then gives 
\be
\label{tvtau1}
t-t_* =
{2\over 2x_i+3\gamma y_i} 
\sqrt{{(3+2\omega)\Phi_0a_0^{3\gamma} \over 8\pi \Gamma}}
\ e^{(2x_i+3\gamma y_i)\tau/2} \ .
\ee
This ensures that we can only have cosmic time $t\to\infty$ at
critical points in the finite phase-plane when
$\tau\to\infty$ (for $2x_i+3\gamma y_i\geq0$) and $t\to-\infty$ when
$\tau\to-\infty$ (for $2x_i+3\gamma y_i\leq0$).
Critical points thus correspond to power-law solutions
with respect to cosmic time for the scale factor $a$ and scalar field
$\Phi$ of the form given in Eq.~(\ref{powerlaw}) where
\be
\label{PLSOLNS}
p_i = \frac{2y_i}{2x_i + 3\gamma y_i} \ , \qquad
q_i = \frac{4x_i}{2x_i + 3\gamma y_i} \ .
\ee

\subsubsection{Point $(a)$}

Critical point $(x_a,y_a)$ always lies on the $k=0$ separatrix, as can
be seen by comparing Eqs.~(\ref{SOLA}) and~(\ref{EXPLICITK}). 
Substituting Eq.~(\ref{XAYA}) into Eq.~(\ref{PLSOLNS}) we see that
point $(a)$ corresponds to Nariai's power-law solution for $k=0$ given
in Eqs.~(\ref{NARIAIp}) and~(\ref{NARIAIq}).

\begin{figure}[t]
\centering 
\leavevmode\epsfysize=10cm \epsfbox{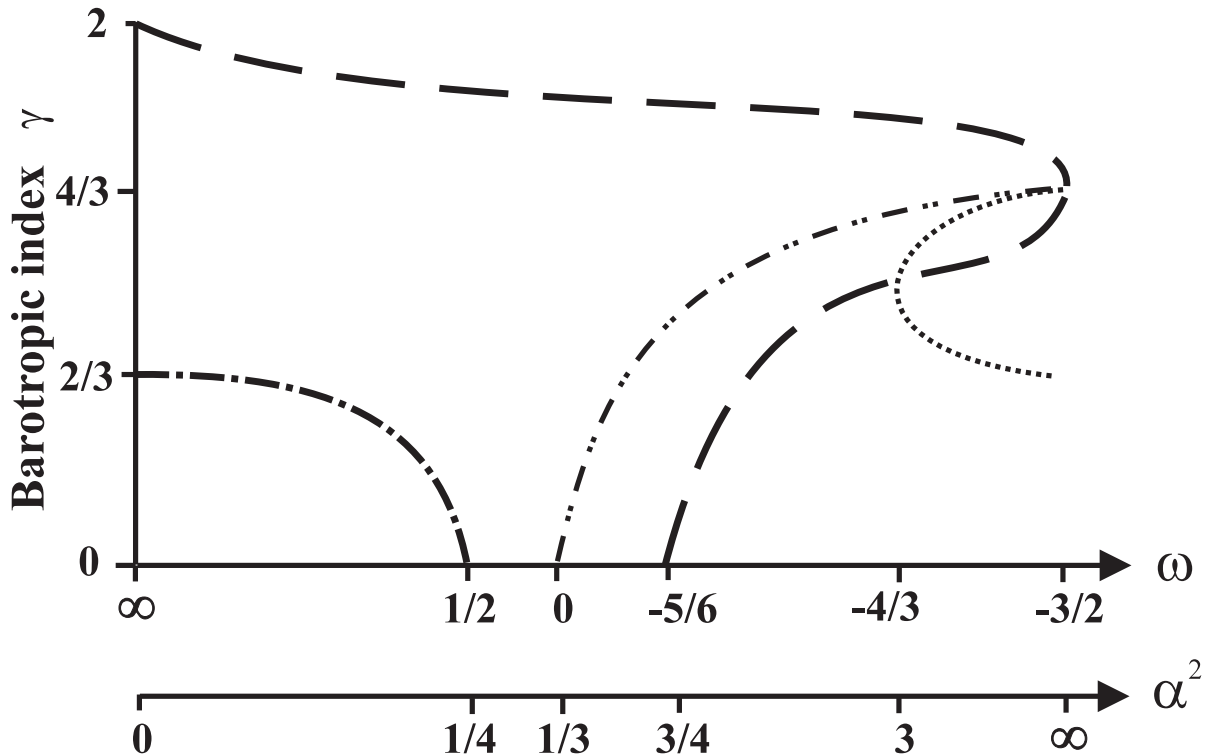}\\ 
\caption[Plane Divisions]{\label{PlaneDivisions} 
Lines dividing regions in parameter space with different cosmological
behaviour at critical point $(a)$.  The dashed line represents
$\omega_a(\gamma)$, defined in Eq.~(\ref{OMEGAA}). The single
dot-dashed line represents $\omega_*(\gamma)$, defined in
Eq.~(\ref{OMEGA*}).  The double dot-dashed line represents
$\omega_+(\gamma)$, defined in Eq.~(\ref{OMEGA+}). The dotted line
represents $\omega_-(\gamma)$ as defined in Eq.~(\ref{OMEGA-}).}
\end{figure}

{}From Eq.~(\ref{XAYA}) we see that point $(a)$ exists for
$\alpha<\sqrt{3}(2-\gamma)/|4-3\gamma|$ which implies that
$\omega>\omega_a(\gamma)$, where we define
\be
\label{OMEGAA}
\omega_a(\gamma) \equiv \frac{-(10-6\gamma)}{3(2-\gamma)^2}
\ee
It can be seen that $(x_a,y_a)$ tends to infinity as
$\omega\to\omega_a(\gamma)$. In the limit $\gamma\to2$ we require
$\omega\to\infty$ for $(x_a,y_a)$ to exist at finite values in the
phase plane. On the other hand, as $\omega\to-3/2$ we find that
$\gamma$ is restricted to $4/3$ for point $(a)$ to exist, as shown in
Figure~\ref{PlaneDivisions}. 

For point $(a)$ to correspond to an expanding universe, $\dot{a}>0$,
requires $y_a>0$ and hence, from Eq.~(\ref{XAYA}),
$\omega>\omega_+(\gamma)$, where we define
\be
\label{OMEGA+}
\omega_+(\gamma) \equiv {-1 \over 2-\gamma}
\ee
Note also that we have $t\to+\infty$ as $\tau\to+\infty$ in
Eq.~(\ref{tvtau1}) at point $(a)$ only if
$\omega\geq\omega_-(\gamma)$, where we define
\be
\label{OMEGA-}
\omega_-(\gamma) \equiv \frac{-4}{3\gamma(2-\gamma)} \ .
\ee
Conversely we have $t\to-\infty$ as $\tau\to-\infty$ only for
$\omega\leq\omega_-(\gamma)$. In the particular case
$\omega=\omega_-(\gamma)$, point $(a)$ corresponds to a non-singular
exponential expansion, $a\propto\exp(Ht)$, where the Hubble rate, $H$,
is a constant.

Nariai's solution corresponds to an accelerated scale factor,
$\ddot{a}>0$, if $p>1$ or $p<0$. 
For $p<0$ this requires $\omega_-(\gamma)<\omega<\omega_+(\gamma)$,
in which case point $(a)$ corresponds to a contracting universe.
For $p>1$ we require $\omega>\omega_*(\gamma)$ when $\gamma<2/3$, 
where we define
\be
\label{OMEGA*}
\omega_*(\gamma) \equiv \frac{2}{(2-\gamma)(2-3\gamma)} \ ,
\ee
or $\omega_a(\gamma) < \omega \leq \omega_-(\gamma)$ when
$1<\gamma<4/3$. But only in the former case do we have
$\omega>\omega_+(\gamma)$ and hence an accelerated, expanding
universe. See Figure~\ref{PlaneDivisions}.

For the particular case $\gamma=0$ (corresponding to a false vacuum
energy density), point $(a)$ corresponds to
\be
a \propto t^{\omega + 1/2} \ , \qquad \Phi \propto t^2
\ee
which is the basis of models of extended inflation
solution~\cite{extinf} when $\omega>\omega_*(0)=1/2$.

\subsubsection{Point $(b)$}

Critical point $(x_b,y_b)$ is a novel power-law solution which
corresponds to a curvature scaling solution~\cite{WANDS93,Kolitch}.
Substituting Eq.~(\ref{XBYB}) into Eqs.~(\ref{powerlaw})
and~(\ref{PLSOLNS}) we find
\be
a \propto t \ , \qquad \Phi \propto t^{2-3/\gamma} \ .
\ee
Thus we have $\Phi \propto a^{2-3\gamma}$ and so
the gravitational effect of the matter remains proportional
to that of the curvature in the constraint Eq.~(\ref{STCONSTRAINT})
\be
\left(\frac{\dot{a}}{a}\right)^2 \propto \frac{\rho}{\Phi}
 \propto \frac{k}{a^2} \ .
\ee
The universe expands with $a\propto t$, as it would in general
relativity when $\gamma=2/3$ and $\rho\propto1/a^2$ in general
relativity. In Brans-Dicke theory this solution exists for all
$\gamma\leq2/3$.  Point $(b)$ maybe described as marginal inflation,
in the sense that although the spatial curvature does not vanish with
respect to the gravitational energy density in the constraint
equation~(\ref{STCONSTRAINT}), it does not grow either, and remains
proportional to the gravitational energy density as $t\to\infty$.

Point $(b)$ exists for $\gamma \leq 2/3$, irrespective of the value of
$\omega$~\cite{WANDS93,Kolitch}. In the limit $\gamma\to 2/3$ we have
$x_b\to0$ and $y_b\to\infty$. Note that $2x_b+3\gamma y_b>0$ whenever
point $(b)$ exists and thus we have $t\to+\infty$ as $\tau\to+\infty$
in Eq.~(\ref{tvtau1}).

When $\omega=\omega_*(\gamma)$, defined in Eq.~(\ref{OMEGA*}), point
$b$ lies on the $k=0$ separatrix and the critical points $(x_a,y_a)$
and $(x_b,y_b)$ coincide.  When $\omega>\omega_*$ point $(b)$ lies in
the $k>0$ region, while for $\omega<\omega_*$ it lies in the $k<0$
region.

\subsection{Stability of critical points in the finite plane}

To determine the stability of these solutions we expand about the
critical points
\bea
x & = & x_i + u \\
y & = & y_i + v
\eea
and substitute into the equations of motion for $x$ and $y$,
Eqs.~(\ref{XPRIME}) and~(\ref{YPRIME}). 
This yields, to first-order in $u$ and $v$,
\bea
u' & = & - \left( 2 x_i + {3(2-\gamma)\over2} y_i \right) u -
\left({3(2-\gamma) \over 2} x_i\right) v \\
v' &=& \left( - {4(1-3\alpha^2) \over 3\alpha^2} x_i + 3 y_i \right) u
 + \left( 3x_i + (3\gamma-2) y_i \right) v
\eea

We seek the two eigenvalues $\lambda_\pm$ such that
\be
u' = \lambda_\pm u \ , \qquad v' = \lambda_\pm v \ ,
\ee
for the eigenmodes $u=K_\pm v$, where $K_\pm$ are constants.
The eigenvalues are given by the roots of the quadratic
\be
\lambda^2 + B\lambda + C = 0
\ee
where
\bea
B_i & \equiv & {10-9\gamma \over 2} y_i - x_i \\
C_i & \equiv &
 - 6 \left( {(1-3\alpha^2)(2-\gamma) \over 3\alpha^2} + 1 \right) x_i^2
 - 2(3\gamma-2) x_i y_i
 - {3(2-\gamma)(3\gamma-2) \over 2}y_i^2
\eea

The critical point is a stable, attractor solution if the real part of
both the eigenvalues is negative (i.e., $B_i>0$ and $C_i>0$).
Linear perturbations about such points are then exponentially damped. A
saddle point in the phase--plane has one negative and one positive
real root ($C_i<0$).  An unstable node has eigenvalues with
positive real parts ($B_i<0$ and $C_i>0$) corresponding to growing
perturbations.

For the critical point $(b)$ we find that $B_b>0$ for all values of
$\gamma<2/3$ where $(b)$ exists in the finite plane. $C_b$ is positive
for $\omega<\omega_*$ defined in Eq.~(\ref{OMEGA*}) and point $(b)$ is
then a stable attractor~\cite{WANDS93,Kolitch}. For $\omega>\omega_*$
we find $C_b<0$ and point $(b)$ is then a saddle
point~\cite{WANDS93,Kolitch}.

For point $(a)$ we find $C_a>0$ for $\omega>\omega_*$. We also have
$B_a>0$ in this region and thus the point $(a)$ is a stable attractor.
For $\omega<\omega_*$ we have $C_a<0$ and the critical point $(a)$ is
a saddle point.  However point $(a)$ is always the attractor for
solutions on the k=0 separatrix when it exists.

For the case $\omega=\omega_*$ the equations are degenerate and the
critical points $(a)$ and $(b)$ coincident. Further analysis is
required to determine the nature of the critical point. It can be
shown that for $k\le0$ the critical point remains an attractor, whilst
for $k>0$ the critical point is a saddle point. In the phase plane as
a whole the point is a col--node.

\subsection{Critical Points at Infinity}

In order to fully describe the qualitative evolution of the system, we
must understand the nature of the phase plane at infinity. This
corresponds to the regime where the dynamical effect of the fluid
density becomes negligible either with respect to the expansion
($y\to\pm\infty$) or the evolution of the Brans-Dicke field
($x\to\pm\infty$).

Compactification can be achieved by projecting the
infinite $(x,y)$ plane onto the unit disc $(r,\theta)$ where
\be
x = R \cos \theta \ , \qquad
y = R \sin \theta \ , \qquad
r = \frac{R}{1+R} \ .
\ee
The equations of motion for $r$ and $\theta$ then reduce to 
\bea
r\theta' &=& (1-r) \left[ \cos\theta\, y' - \sin\theta\, x' \right] \ ,\\
r' &=& (1-r)^2 \left[ \cos\theta\, x' + \sin\theta\, y' \right] \ ,
\eea
where $x'$ and $y'$ are given by Eqs.~(\ref{XPRIME}) and~(\ref{YPRIME}).

At infinity (the limit $r\to1$) we find
\bea
\label{INFINTHETA}
\theta' &\to& {2\over3\alpha^2(1-r)} \, \cos\theta \,
 \left[ 3\alpha^2 (\sin\theta + \cos\theta)^2 - \cos^2\theta \right] \\
\label{INFINR}
r' &\to& f(\theta) \equiv 
-\cos\theta 
+\left[ 2\left(\sin\theta+\cos\theta\right)^2
        - {2\over3\alpha^2}\cos^2\theta - {3(2-\gamma)\over2} \right]
\sin\theta \ .
\eea

The equation for $\theta'$ determines the behaviour of the solutions
at infinity ($r=1$) and is independent of $\gamma$.  Solving for
$\theta'=0$ as $r\to1$ we find the critical values $\theta=\theta_i$,
where
\be
\label{TANTHETAI}
\tan\theta_c = -1 + {1\over\sqrt{3}\alpha} \ , \quad
\tan\theta_d = -1 - {1\over\sqrt{3}\alpha} \ , \quad
\cos\theta_e = 0 \ .
\ee
Clearly there are two solutions for each case in the complete interval
$(0,2\pi)$. For each solution $\theta_{i+}$ there is a solution
$\theta_{i-}$ which is related by $\theta_{i-}=\theta_{i+}+\pi$ and
corresponds to the time reverse of the original solution. Just as we
did for the critical points at finite values of $x$ and $y$, in the
rest of this section we will consider only the solutions $\theta_{i+}$
which lie in the interval $(-\pi/4,+3\pi/4)$ corresponding to
$x_i+y_i>0$. 

Note that $|d\theta/dr|\to\infty$ whenever $\theta'\neq0$ as $r\to1$
and thus all trajectories (except those with $r=1$ at all times)
approach the critical points at infinity along a radial
trajectory. Approaching infinity with $\theta=\theta_i$ at a finite
value of $\tau\to\tau_*$ we have, from Eqs.~(\ref{XPRIME}),
(\ref{YPRIME}) and~(\ref{INFINR}),
\be
\Phi \propto |\tau-\tau_*|^{-2\cos\theta_i/f(\theta_i)} \ , \qquad
a\propto|\tau-\tau_*|^{-\sin\theta_i/f(\theta_i)} \ .
\ee
Integration of Eq.~(\ref{DTAU}) then gives
\be
\label{tvtau2}
t-t_* \propto 
\left| \tau-\tau_* \right|^{-F(\theta_i)/f(\theta_i)} \ ,
\ee
where
\bea
\label{bigF}
F(\theta_i) & \equiv &
{2\cos\theta_i+3\gamma\sin\theta_i-2f(\theta_i) \over 2} \nonumber \\
~& = & 2\cos\theta_i + \left[ 3 - 2\left(\sin\theta_i+\cos\theta_i\right)^2
+ {2\over3\alpha^2} \cos^2\theta_i \right] \sin\theta_i
\ . 
\eea
Thus the critical points at infinity also correspond to
power-law evolution of the scale factor and scalar field with respect
to the proper cosmic time $t$ of the form given in
Eqs.~(\ref{PLSOLNS}) with
\be
\label{INFINpq}
p = {\sin\theta_i \over F(\theta_i)} \ , \qquad
q = {2\cos\theta_i \over F(\theta_i)} \ .
\ee

We find $t\to+\infty$ as $\tau\to\tau_*$ in Eq.~(\ref{tvtau2}) for
$F(\theta_i)>0$ and $f(\theta_i)>0$ or $t\to-\infty$ as
$\tau\to\tau_*$ for $F(\theta_i)<0$ and $f(\theta_i)<0$.

\subsubsection{Points $(c)$ and $(d)$}

Points $(c)$ and $(d)$ lie at either end of the k=0 separatrix,
defined by Eq.~(\ref{k0}).  Substitution of
$\tan\theta_i=-1\pm1/\sqrt{3}\alpha$ into Eq.~(\ref{INFINpq}) gives
the exponents given in Eq.~(\ref{OHpq}) for the power-law evolution of
$a$ and $\Phi$ with respect to the proper time $t$.  These are
spatially flat ($k=0$) vacuum solutions, first given by O'Hanlon and
Tupper~\cite{OHT72}. In terms of the analysis by Gurevich et
al~\cite{GFR73}, points $(c)$ and $(d)$ correspond to the {\it slow} and
{\it fast} solutions respectively.

Point $(d)$ is always an expanding solution ($\sin\theta>0$), but
point $(c)$ corresponds to a contracting solution if $\omega<0$.

\subsubsection{Point $(e)$}

Point $(e)$ lies on the x-axis as $y\to\infty$ in the $k<0$ region.
Substituting $\cos\theta_e=0$ into Eq.~(\ref{INFINpq}) 
gives the evolution of the scale factor and scalar field
with repect to proper time $t$,
\be
\Phi\to {\rm constant} \ , \qquad a \propto t
\ee
Point $(e)$ is therefore the general relativistic curvature--dominated
Milne universe~\cite{Kolitch}.

\subsection{Stability of points at infinity}

By inspecting Eq.~(\ref{INFINTHETA}) for $\theta'$ we see that solutions
at infinity ($r=1$) approach point $(e)$ but diverge away from points
$(c)$ and $(d)$ independently of the parameters $\gamma$ and
$\omega$. The overall stability of the points depends upon the sign of
$r'=f(\theta_i)$ approaching these points from $r<1$.

For point $(e)$ we have $\cos\theta_e=0$ and so from
Eq.~(\ref{INFINR}) we have $f(\theta_e)=(3\gamma-2)/2$. Thus point
$(e)$ is a stable attractor for $\gamma>2/3$. We also find
$F(\theta_e)=1$ in Eq.~(\ref{tvtau2}) and thus $t\to+\infty$ as
$\tau\to\tau_*$ at point $(e)$ whenever $\gamma>2/3$~\cite{Kolitch}.

For points $(c)$ and $(d)$ we can eliminate $1/3\alpha^2$ from
Eq.~(\ref{INFINR}) by the subsitution of $(1+\tan\theta_i)^2$
using Eqn.~({\ref{TANTHETAI}), and we obtain

\be
f(\theta_i) = - {3(2-\gamma) \over 2} \sin\theta_i - \cos\theta_i
\ee
We find that whenever the critical point $(a)$ exists in the finite
phase plane [$\omega>\omega_a(\gamma)$ defined in Eq.~(\ref{OMEGAA})]
we have $f(\theta_i)<0$ for both points $(c)$ and $(d)$ so that the
critical points are unstable nodes.  When point $(a)$ no longer exists
the behaviour at the critical point depends on the value of $\gamma$.
For $\gamma<4/3$ point $(c)$ is a saddle point but is an attractor for
$k=0$ models, whilst $(d)$ is an unstable node. For $\gamma>4/3$ the
converse is true.

Note also that for points $(c)$ and $(d)$ we find $F(\theta_i)=
2\cos\theta_i+3\sin\theta_i$ in Eq.~(\ref{tvtau2}), we have
$F(\theta_d)>0$ for all $\omega$, whilst $F(\theta_c)>0$ for
$\omega>-4/3$. Thus for point $(c)$ we have $t\to+\infty$ as
$\tau\to\tau_*$ for $-4/3<\omega<\omega_a(\gamma)$ (when $\gamma<1$),
and conversely $t\to-\infty$ as $\tau\to\tau_*$ for $\omega<-4/3$
(when $\gamma>4/3$) or for $\omega_a(\gamma)<\omega<-4/3$ (when
$1<\gamma<4/3$).  For point $(d)$ $t\to+\infty$ as $\tau\to\tau_*$ for
$\omega<\omega_a(\gamma)$ (when $\gamma>4/3$), there are no regions of
the phase plane where $t\to-\infty$ as $\tau\to\tau_*$.

\section{Qualitative evolution}
\label{sectqual}
\setcounter{equation}{0}

In this section we present a qualitative description for the dynamical
evolution of the phase plane for different regions of the
$(\omega,\gamma)$ parameter space. See Figure~\ref{regions}.

\begin{figure}[t]
\centering 
\leavevmode\epsfysize=10cm \epsfbox{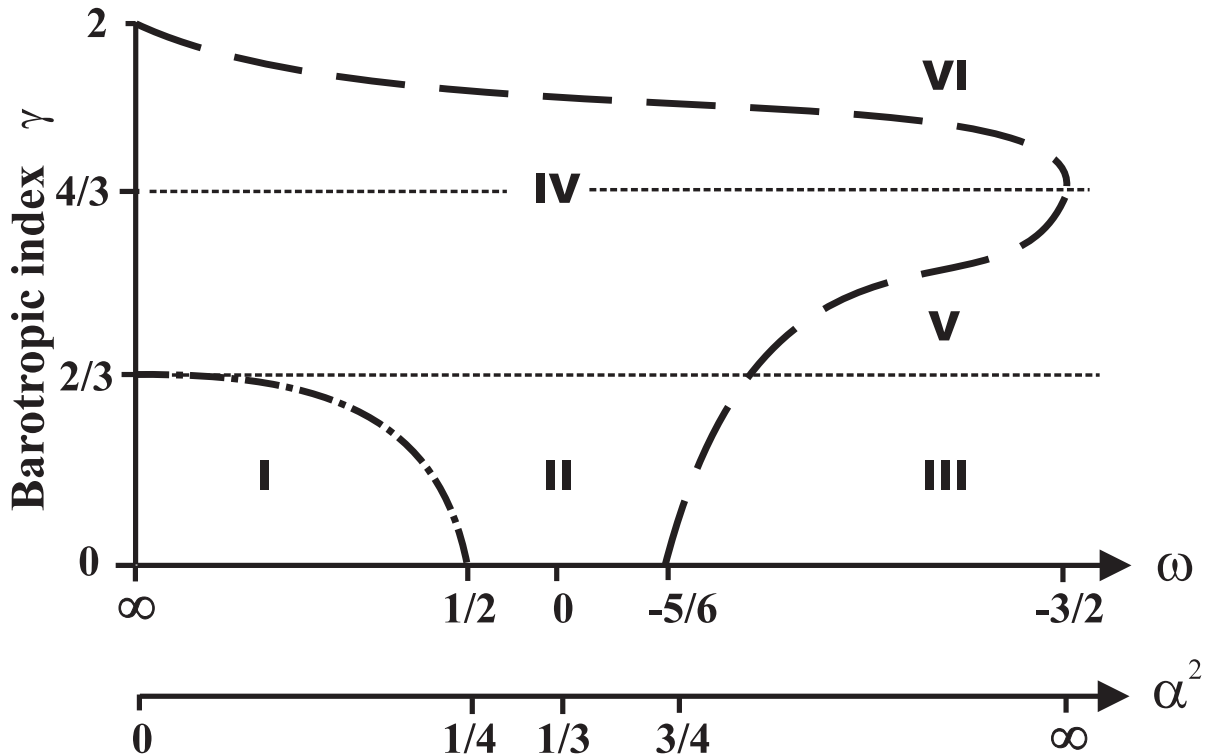}\\ 
\caption[regions]{\label{regions} Parameter regimes defined in
Section~\ref{sectqual}. 
The dashed line represents $\omega_a(\gamma)$ and the single dot-dashed
line represents $\omega_*(\gamma)$.}
\end{figure}

\subsection{Region I: \qquad $\omega_*(\gamma) < \omega <
\infty \ , \ \gamma < 2/3$.~~~{\rm See Figure~\ref{plot1}.}}

Critical points $(a_\pm)$ and $(b_\pm)$ exist at finite values of $x$ and
$y$, with point $(a_+)$ being the stable late-time attractor on the
$x+y>0$, $k=0$ separatrix. This parameter regime is inflationary as
point $(a_+)$ corresponds to accelerated expansion and is the attractor
solution for all $k\leq0$ models with $x+y>0$ and a non-zero measure
of the $k>0$ models. The remaining $k>0$ trajectories and all $k\leq0$
models with $x+y<0$ approach one of the vacuum solutions $(c_-)$ or
$(d_-)$.
We find that only those closed ($k>0$) models above a second sepratrix
passing through point $(b_+)$ reach the inflationary solution at late
times~\cite{WANDS93,Kolitch}.
For instance, when $\omega=18(1-\gamma)/(2-3\gamma)^2$ the second
separatrix is the straight line $y=2x/(2-\gamma)$, but we do not have
an analytic expression for this separatrix in the more general case.

The generic early-time behaviour is given by the matter dominated
solution $a_-$ or the vacuum solutions $c_+$ or $d_+$, while points
$b_\pm$ and $e_\pm$ are saddle points.

\begin{figure}[t]
\centering 
\leavevmode\epsfysize=10cm \epsfbox{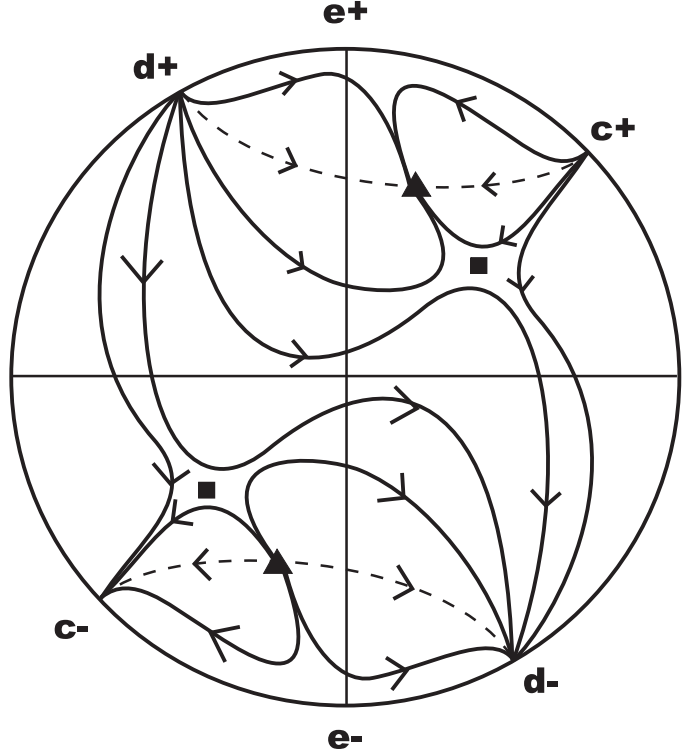}\\ 
\caption[plot1]{\label{plot1} 
Example of phase-plane in region~I ($\gamma=1/3, \omega=6$). 
The solid triangle represents critical points $(a_\pm)$. The solid
square represents the critical points $(b_\pm)$. The dashed line
is the $k=0$ separatrix.}
\end{figure}

\subsection{Region II: \qquad $\omega_a(\gamma)< \omega < \omega_*(\gamma) 
\ , \ \gamma < 2/3 $.~~~{\rm See Figure~\ref{plot2}.}}

Critical points $a_\pm$ and $b_\pm$ exist at finite values of $x$ and
$y$. We have marginal inflation as point $b_+$ the stable late-time
attractor solution for all $k<0$ models with $x+y>0$. All $k>0$
trajectories and $k\leq0$ models with $x+y<0$ collapse to the vacuum
solutions $c_-$ or $d_-$.

The generic early-time behaviour is given by the scaling solution
$b_-$ for $k<0$ models with $x+y<0$ or the vacuum solutions $c_+$ or
$d_+$ for $k\leq0$ trajectories with $x+y>0$ and all $k>0$ models.
Points $a_\pm$ and $e_\pm$ are saddle points.

As $\omega\to\omega_a$ from above (and we approach region III) point
$(a_+)$ moves along the $k=0$ separatrix towards point $(c_+)$ at
infinity, and $(a_-)$ moves off towards $(d_-)$. On the other hand, as
$\gamma\to2/3$ from below (and we approach region IV) we see that
points $(b_\pm)$ move off towards the curvature dominated solutions at
points $(e_\pm)$.

\begin{figure}[t]
\centering 
\leavevmode\epsfysize=10cm \epsfbox{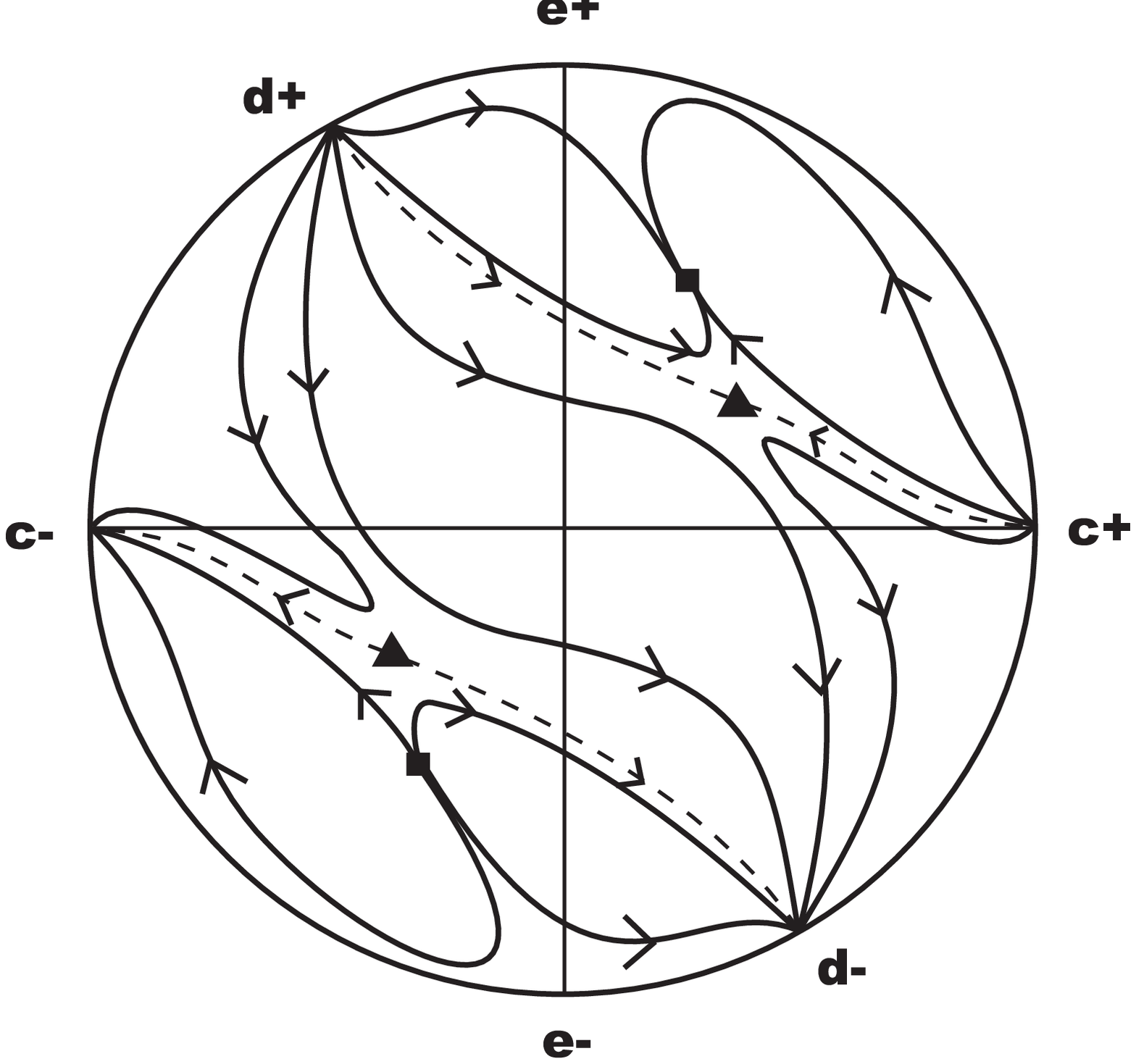}\\ 
\caption[plot2]{\label{plot2}
Example of phase-plane in region~II ($\gamma=1/3, \omega=0$).
Symbols as for Figure~\ref{plot1}.}
\end{figure}

\subsection{Region III: \qquad $\omega < \omega_a(\gamma) \ , 
\ \gamma < 2/3 $.~~~{\rm See Figure~\ref{plot3}.}}

Only points $b_\pm$ now exist at finite values of $x$ and $y$. We have
marginal inflation as point $b_+$ is the stable late-time attractor
for all $k<0$ models with $x+y>0$. All $k>0$ models and $k\leq0$
trajectories with $x+y<0$ approach the vacuum solution $d_-$ at late
times.  Generic $k=0$ solutions with $x+y>0$ evolve from $d_+$ to
$c_+$ while $k=0$ solutions with $x+y<0$ evolve from $c_-$ to $d_-$.

The generic early-time behaviour is given by the scaling solution
$b_-$ for $k<0$ models with $x+y<0$ or the vacuum solution $d_+$ for
$k\leq0$ trajectories with $x+y>0$ and all $k>0$ models.
Points $c_\pm$ and $e_\pm$ are saddle points.

As $\gamma\to2/3$ from below (and we approach region VI) points
$(b_\pm)$ tend towards points $(e_\pm)$.

\begin{figure}[t]
\centering 
\leavevmode\epsfysize=10cm \epsfbox{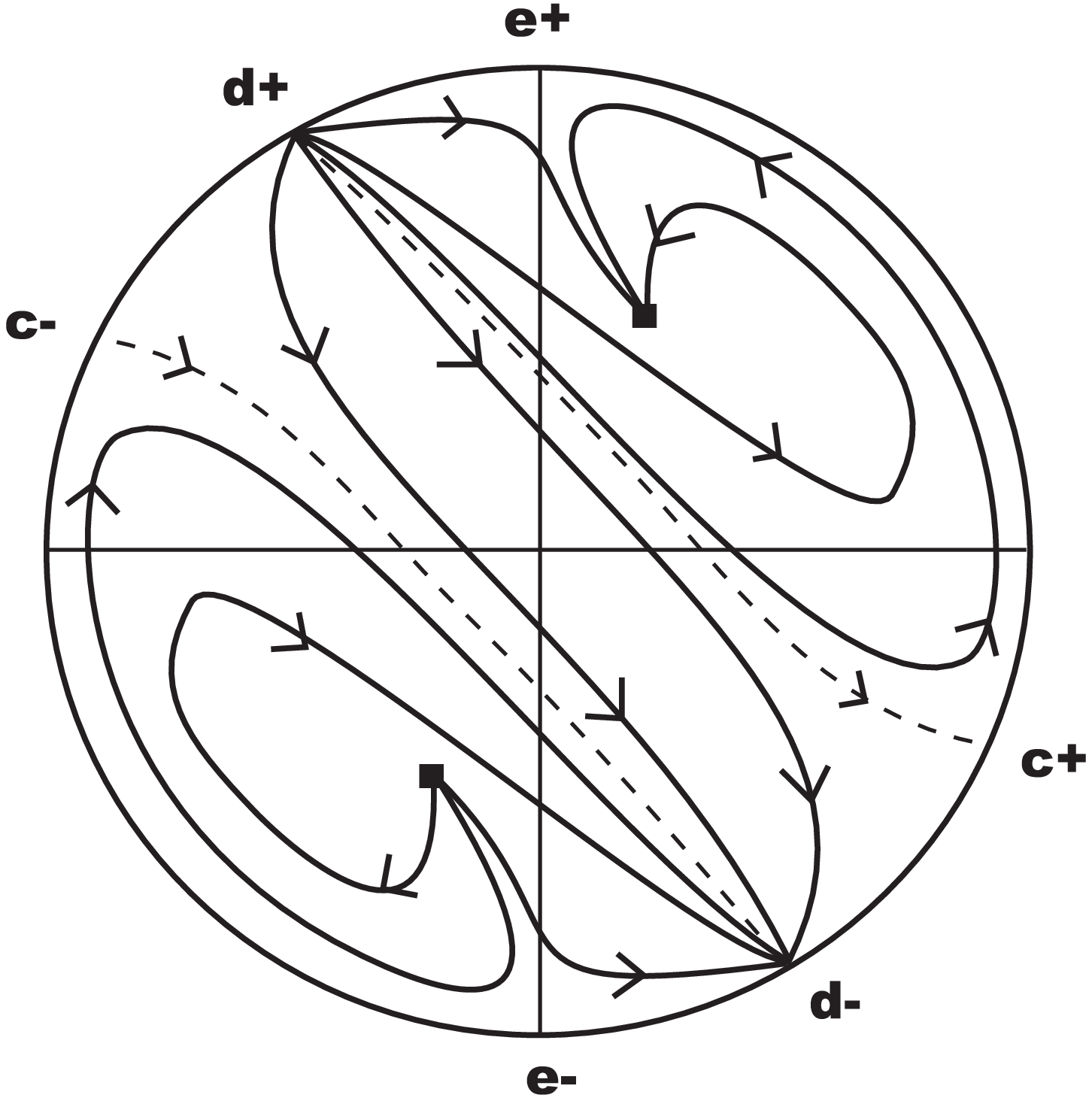}\\ 
\caption[plot3]{\label{plot3} 
Example of phase-plane in region~III ($\gamma=1/3, \omega=-1$).
Symbols as for Figure~\ref{plot1}.}
\end{figure}

\subsection{Region IV: \qquad $\omega_a(\gamma) < \omega <
\infty \ , \ \frac{2}{3} < \gamma < 2$.~~~{\rm See Figure~\ref{plot4a}
and~\ref{plot4b}.}} 

Only points $a_\pm$ exist at finite values of $x$ and $y$. Point $e_+$
is the stable late time attractor for $k<0$ solutions with
$x+y>0$. All $k>0$ models and $k\leq0$ trajectories with $x+y<0$
approach one of the vacuum solutions $c_-$ or $d_-$ at late times.
Generic $k=0$ solutions with $x+y>0$ evolve from vacuum solutions
$c_+$ or $d_+$ towards the matter dominated solution $a_+$.

The generic early-time behaviour is given by point $e_-$ for $k<0$
models with $x+y<0$ or the vacuum solutions $c_+$ or $d_+$ for
$k\leq0$ trajectories with $x+y>0$ and all $k>0$ models.
Points $a_\pm$ are saddle points.

For $\gamma<4/3$ we see that as $\omega\to\omega_a$ from above (and we
approach region V) point $(a_+)$ moves off towards point $(c_+)$, and
$(a_-)$ moves off towards $(d_-)$ at infinity. On the other hand, as
$\omega\to\omega_a$ from above for $\gamma>4/3$ (and we approach
region VI) point $(a_+)$ moves off towards point $(d_+)$, and $(a_-)$
moves off towards $(c_-)$ at infinity.

\begin{figure}[t]
\centering 
\leavevmode\epsfysize=10cm \epsfbox{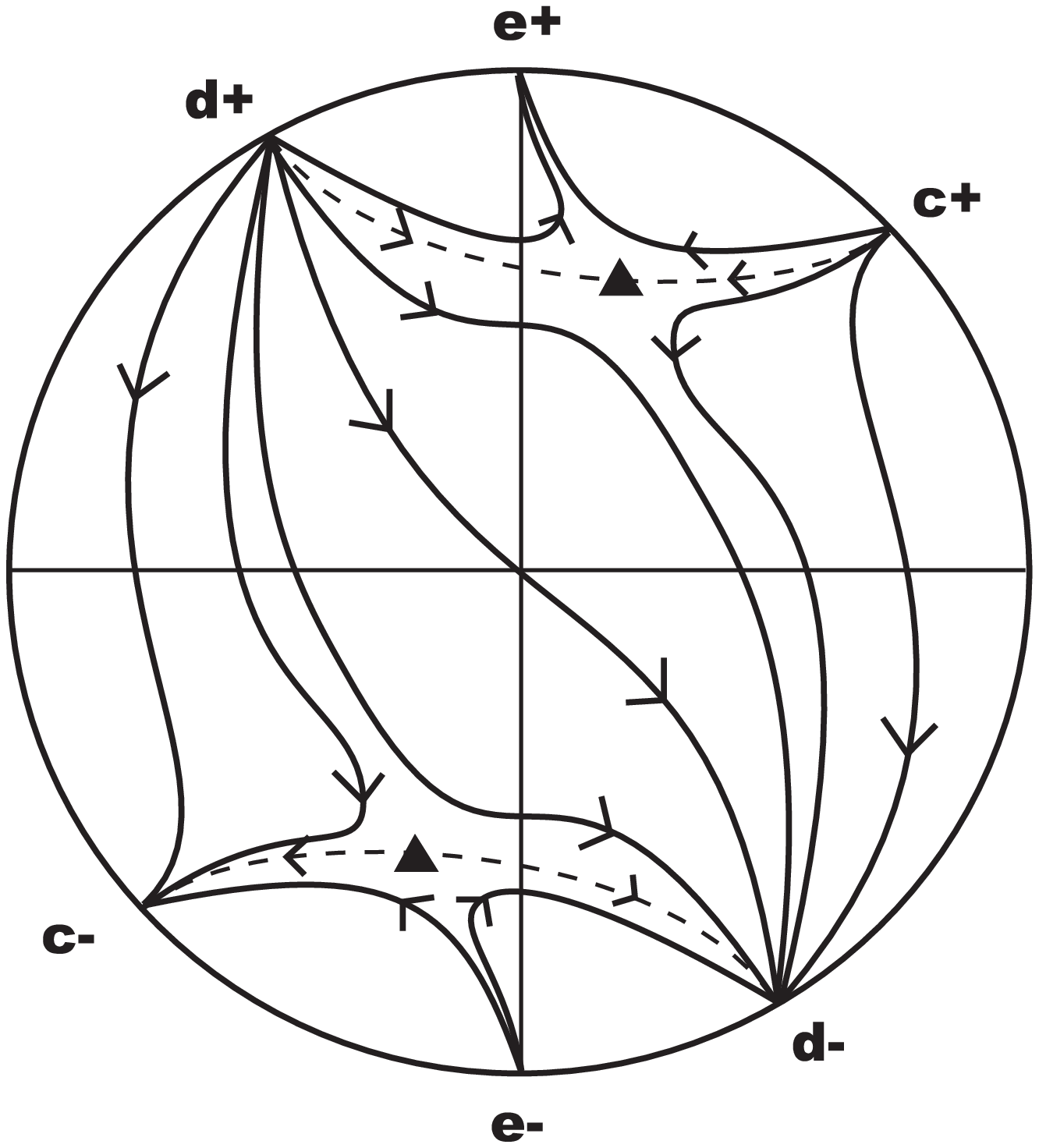}\\ 
\caption[plot4a]{\label{plot4a} 
Example of phase-plane in region~IV when $\gamma<4.3$ 
(here $\gamma=1, \omega=6$).
Symbols as for Figure~\ref{plot1}.}
\end{figure}
\begin{figure}[t]
\centering 
\leavevmode\epsfysize=10cm \epsfbox{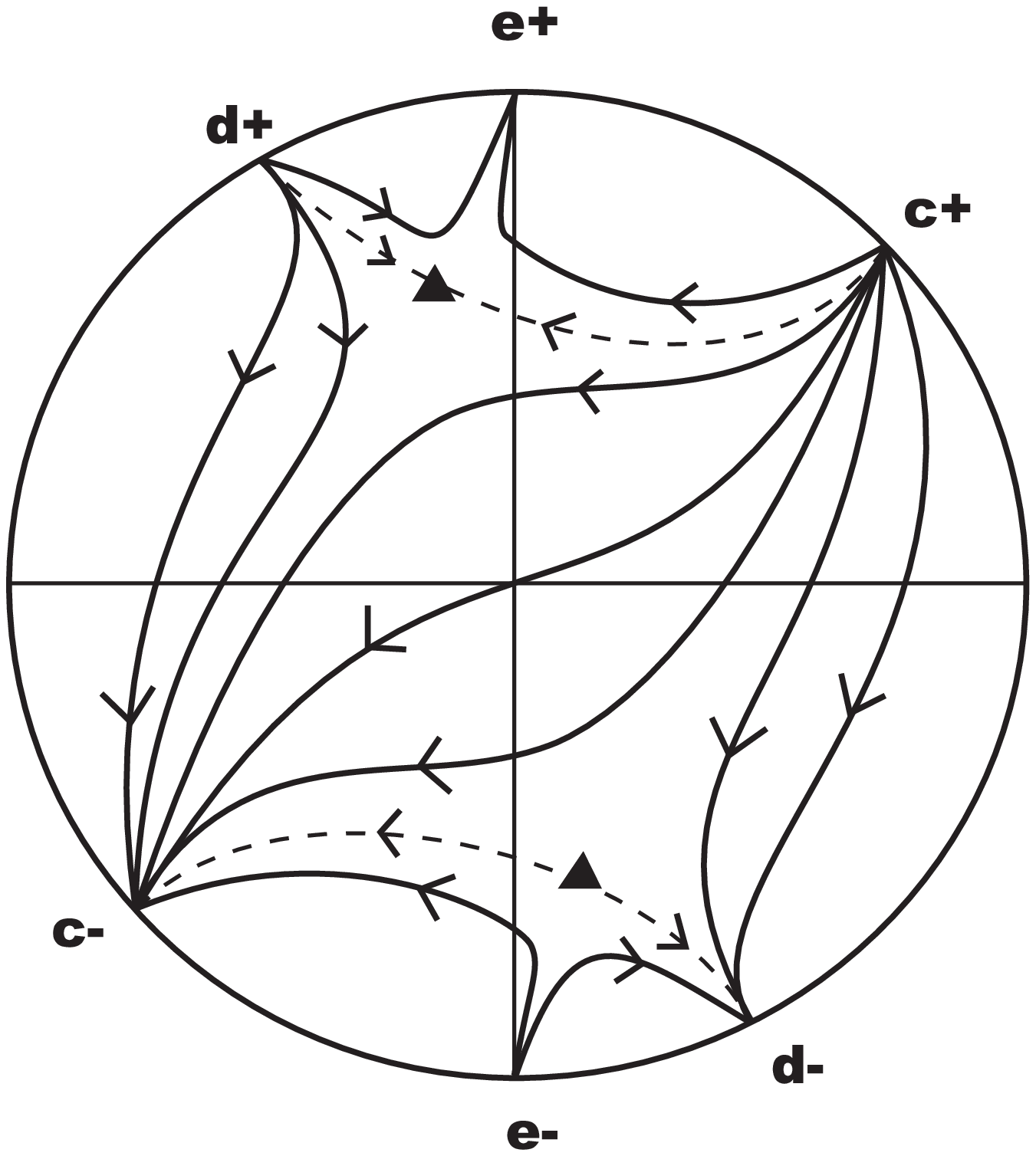}\\ 
\caption[plot4b]{\label{plot4b} 
Example of phase-plane in region~IV when $\gamma>4.3$ 
(here $\gamma=1.5, \omega=6$).
Symbols as for Figure~\ref{plot1}.}
\end{figure}

\subsection{Region V: \qquad $\omega < \omega_a(\gamma)
\ , 2/3 < \gamma < 4/3$.~~~{\rm See Figure~\ref{plot5}.}} 

There are no critical points at finite values of $x$ and $y$. Point
$e_+$ is the stable late time attractor for $k<0$ solutions with
$x+y>0$. All $k>0$ models and $k\leq0$ trajectories with $x+y<0$ approach
the vacuum solution $d_-$ at late times.

The generic early-time behaviour is given by point $e_-$ for $k<0$
models with $x+y<0$ or the vacuum solutions $d_+$ for $k\leq0$
trajectories with $x+y>0$ and all $k>0$ models.
Points $c_\pm$ are saddle points.

\begin{figure}[t]
\centering 
\leavevmode\epsfysize=10cm \epsfbox{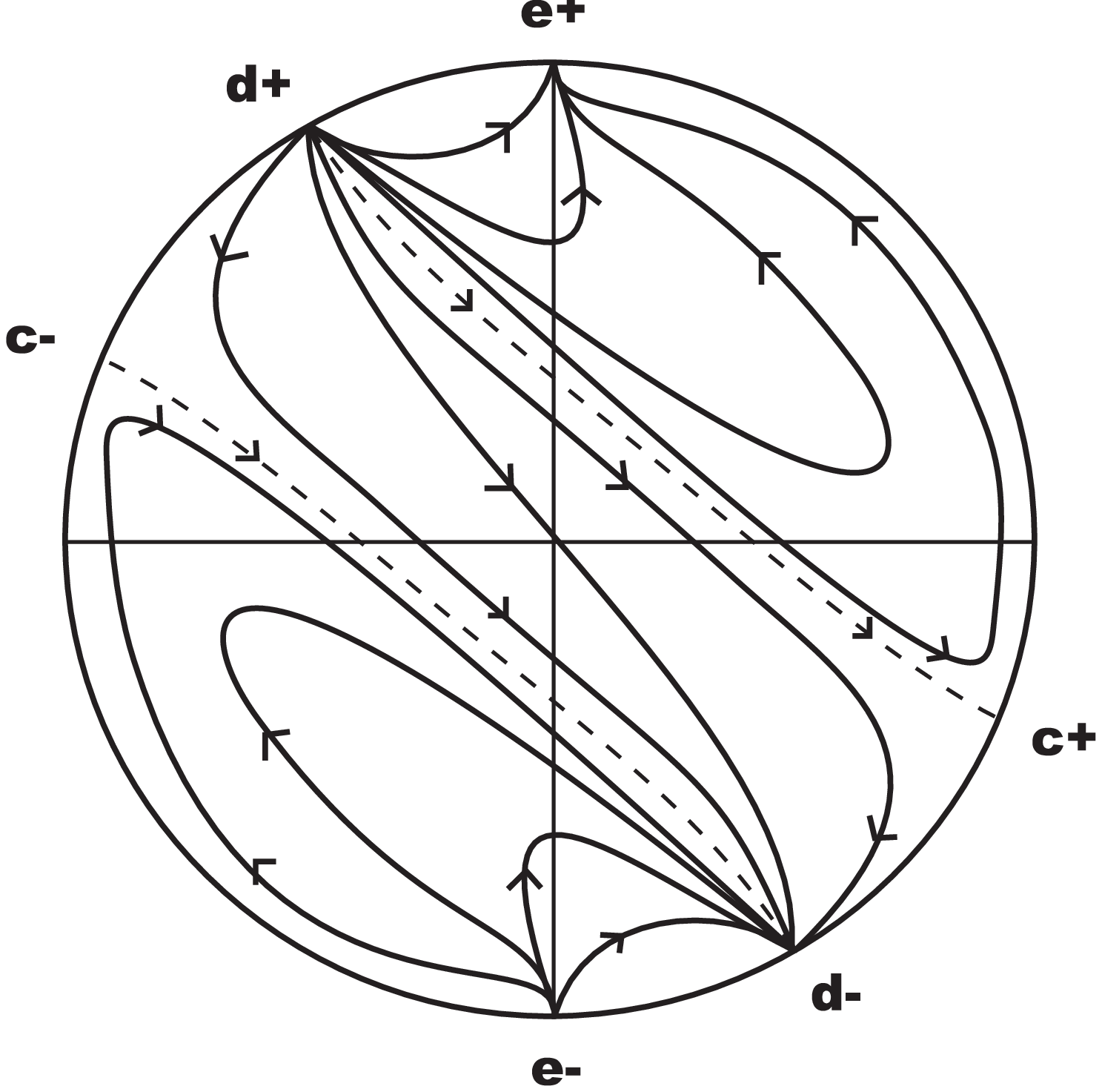}\\ 
\caption[plot5]{\label{plot5} 
Example of phase-plane in region~V ($\gamma=1, \omega=-17/12$).
Symbols as for Figure~\ref{plot1}.}
\end{figure}

\subsection{Region VI: \qquad $\omega < \omega_a(\gamma)
\ , 4/3< \gamma < 2$.~~~{\rm See Figure~\ref{plot6}.}} 

There are no critical points at finite values of $x$ and $y$. Point
$e_+$ is the stable late time attractor for $k<0$ solutions with
$x+y>0$. All $k>0$ models and $k\leq0$ trajectories with $x+y<0$ approach
the vacuum solution $c_-$ at late times.

The generic early-time behaviour is given by point $e_-$ for $k<0$
models with $x+y<0$ or the vacuum solutions $c_+$ for $k\leq0$
trajectories with $x+y>0$ and all $k>0$ models.

Points $d_\pm$ are saddle points.

\begin{figure}[t]
\centering 
\leavevmode\epsfysize=10cm \epsfbox{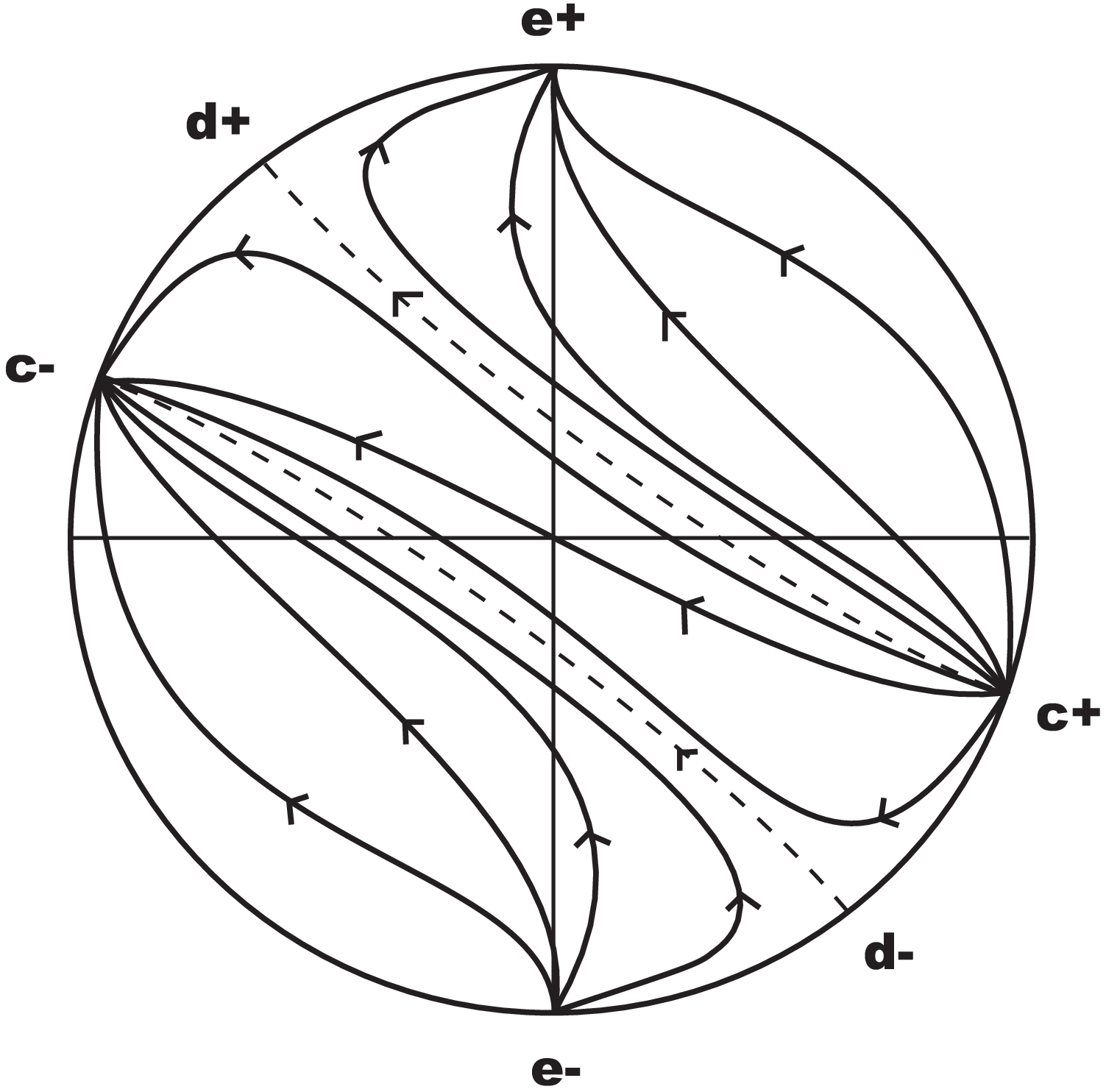}\\ 
\caption[plot6]{\label{plot6}
Example of phase-plane in region~VI ($\gamma=3/2, \omega=-17/12$).
Symbols as for Figure~\ref{plot1}.}
\end{figure}

\section{Non-singular evolution}

In general relativity it is well-known that all spatially flat or open
($k\leq0$) FRW models with a barotropic fluid have a semi-infinite
lifetime and posseess a singularity either at a finite cosmic time in
the past (for expanding models) or in the future (contracting
solutions) so long as $\gamma>0$. Closed models ($k>0$) have a
finite lifetime with both a past and a future singularity unless
$\gamma<2/3$ in which case non-singular evolution, i.e., infinite
proper lifetime, is possible.

It is natural to ask whether in the context of Brans-Dicke gravity it
is possible to have a non-singular universe in open or flat models,
and/or for $\gamma>2/3$. Our phase-plane analysis provides a
qualitative description of the asymptotic evolution that is possible
in the different parameter regimes in terms of the time parameter
$\tau$, and equations~(\ref{tvtau1}) and (\ref{tvtau2}) relate the
asymptotic behaviour of $\tau$ to the proper cosmic time $t$.

As $\omega\to\infty$ the critical points $(a_+)$ and $(b_+)$ with
$x_i+y_i>0$ follow the pattern described above for expanding models in
general relativity with semi-infinite proper lifetimes. Point $(b_+)$,
whenever it exists ($\gamma<2/3$), always describes a solution with a
past singularity, but continuing in to the infinite future. Point
$(a_+)$ similarly describes a semi-infinite lifetime solution with
past singularity for $\omega>{\rm max}\{\omega_a,\omega_-\}$.  Points
$(a_-)$ and $(b_-)$ are just the time-reversed solutions of points
$(a_+)$ and $(b_+)$ respectively, and thus are also semi-infinite but
with the future singularities.  
This suggests that solutions which interpolate between points
$(a/b_-)$ and $(a/b_+)$ would have an infinite lifetime, but the
phase-plane plots show that the solutions which cross the line $x+y=0$
always start from a critical point with $x_i+y_i>0$ and end up at a
critical point $x_j+y_j<0$ which would yield a finite lifetime. These
are analogous to the closed models in general relativity which evolve
from a past singularity to a future singularity.

However for $\omega_a<\omega<\omega_-$ (which is possible for
$1<\gamma<4/3$) the solution at point $(a_+)$ extends to the infinite
past and the singularity lies in the future, whereas it is the point
$(a_-)$ which has a singularity in the past. But in this case we find
no trajectories which link point $(a_+)$ to point $(a_-)$. (Points
$(b_\pm)$ do not exist in this parameter regime.)

Solutions at the fixed points $(a)$ or $(b)$ are necessarily singular
to the past or future with semi-infinite lifetime because they
correspond to power-law evolution for the scale factor, except when
$\omega=\omega_a(\gamma)$. In this particular case the scale factor
expands exponentially and the solutions at points $(a_\pm)$ is indeed 
non-singular. However in this case the solution is unstable as points
$(a_\pm)$ are saddle points in the phase-plane.

Generic solutions in the phase-plane interpolate between two of the
critical points including at least one of the vacuum solutions
$(c_\pm)$, $(d_\pm)$ or $(e_\pm)$ at infinity. As $\omega\to\infty$
points $(c_+)$ and $(d_+)$ are early-time attractors [$f\theta<0$ in
equation~(\ref{INFINR})] as $\tau\to\tau_*$ and this corresponds to a
singularity at finite time $t_*$ [$F(\theta)>0$ in
equation~(\ref{tvtau2})]. For $\gamma<2/3$ point $(e_+)$ is also an
early-time attractor with a singularity at a finite time $t_*$, but
for $\gamma>2/3$ point $(e_+)$ becomes a late-time attractor where
$t\to+\infty$ as $\tau\to\tau_*$. For all values of $\gamma$ we are
unable to construct non-singular trajectories as
$\omega\to\infty$. The situation is actually worse than in general
relativity plus barotropic fluid due to the presence of the scalar
field. Even in the limit $\omega\to\infty$ the Brans-Dicke field leads
to singular evolution at early or late times even in closed models
when $\gamma<2/3$.

However for $\omega_a(\gamma)<\omega<-4/3$, which is possible for
$\gamma>1$, we have $t\to-\infty$ at point $(c_+)$ which is still an
early-time attractor. Thus trajectories in this parameter regime that
originate at point $(c_+)$, which has no past singularity, are
non-singular if they connect to a critical point with no future
singularity. We find three parameter regions in which non-singular
trajectories occur:
\begin{enumerate}
\item
$\omega_a<\omega<\omega_-$
\begin{itemize}
\item
Open models ($k<0$) can originate at point $(c_+)$ in the infinite past
and approach point $(e_+)$ in the infinite future.
\end{itemize}
\item
${\rm max}\{\omega_-,\omega_a\}<\omega<-4/3$
\begin{itemize}
\item
Open models ($k<0$) can originate at point $(c_+)$ in the infinite past
and approach point $(e_+)$ in the infinite future.
\item
Flat models $(k=0$) can originate at point $(c_+)$ in the infinite past
and approach point $(a_+)$ in the infinite future.
\item
Closed models ($k>0$), with $\gamma>4/3$, can originate at point $(c_+)$
in the infinite past and approach point $(c_-)$ in the infinite future.
\end{itemize}
\item
$\omega<{\rm min}\{\omega_-,-4/3\}$
\begin{itemize}
\item
Open models ($k<0$) originate at point $(c_+)$ in the infinite past
and approach point $(e_+)$ in the infinite future.
\item
Flat models $(k=0$) originate at point $(c_+)$ in the infinite past
and approach point $(d_+)$ in the infinite future.
\item
Closed models originate at point $(c_+)$ in the infinite past
and approach point $(c_-)$ in the infinite future.
\end{itemize}
\end{enumerate}
Note that the time reverse of these solutions linking points $(e_-)$
and $(c_-)$, etc., will also be non-singular in the appropriate
parameter regime.

We should stress that an infinite proper lifetime is a necessary, but
not sufficient, condition for these solutions to be non-singular. In
particular, Kaloper and Olive~\cite{Kaloper} have recently emphasised
that as the metric is, by definition, only minimally coupled to the
other fields in the Einstein frame, gravitational waves follow
geodesics in this frame. Thus if the solutions have a singularity in
the finite past in terms of the proper time in the Einstein frame,
defined in Eq.~(\ref{Etime}), they may still be geodesically
incomplete. One can show that all the critical points with $x_i+y_i>0$
have singularities at a finite proper time in the past, and extend
into the infinite future, in the Einstein frame.

\section{Summary}

We have constructed an autonomous phase-plane describing the evolution
of the scale factor and Brans-Dicke field for FRW cosmologies
containing a barotropic fluid in Brans--Dicke gravity. We have
improved upon previous phase-plane analyses~\cite{Kolitch} by
presenting all the FRW models in a single phase-plane.  We are thus
able to present a qualitative analysis of the general evolution of
homogeneous and isotropic cosmologies in Brans-Dicke gravity,
recovering previously known power-law solutions as fixed points in the
phase-plane. These fixed points correspond to self-similar
evolution~\cite{WE97}. This is possible due to the scale-invariance of
the Brans-Dicke gravity theory which contains no characteristic value
for the Brans-Dicke field, $\Phi$.

We find four critical points at finite values of the phase-plane when
$\gamma<2/3$ and $\omega>\omega_a(\gamma)$ [see Eq.~(\ref{OMEGAA})].
Two of the fixed points correspond to Nariai's power-law
solutions~\cite{Nariai68}, $a\propto t^p$, $\Phi\propto t^q$, for
spatially flat models ($k=0$). One of these fixed points is expanding
$y>0$ and the other is contracting $y<0$. The remaining two fixed
points describe a novel self-similar curvature-scaling evolution for
spatially curved models~\cite{WANDS93,Kolitch}, with $a\propto t$ and
$\Phi\propto t^{2-3/\gamma}$.

Nariai's expanding solution is only an attractor at late times if
$\omega>\omega_*(\gamma)$ [see Eq.~(\ref{OMEGA*})]. This corresponds to
power-law inflationary solutions with exponent $p>1$ and a non-zero
measure of spatially curved models approach this flat-space solution. For
instance, for a false vacuum energy density ($\gamma=0$) we have
$\omega_*=1/2$.  On the other hand, for a given value of $\omega$ the
necessary condition for inflation becomes $\gamma<\gamma_*(\omega)$,
where~\cite{BM90,WANDS93}, from Eq.~(\ref{OMEGA*}),
\be
\gamma_* (\omega) \equiv  {2\over 3} \left( 2 - \sqrt{{2\omega+3\over2\omega}}
\right) \ ,
\ee
which is always stronger than in general relativity ($\omega\to\infty$)
where $\gamma_*=2/3$. For $\gamma_*(\omega)<\gamma<2/3$ the expanding
curvature-scaling solution with $k<0$ is a late-time attractor for $k<0$
models.

For $\gamma>2/3$ and $\omega>\omega_a(\gamma)$ only the two critical
points corresponding to Nariai's solutions remain at finite values in
the phase-plane, while for $\gamma<2/3$ but $\omega<\omega_a(\gamma)$
only the curved space fixed points remain. For $\gamma>2/3$ and
$\omega<\omega_a(\gamma)$ their are no fixed points in the finite
phase-plane.

By projected the infinite phase-plane onto a unit disc we studied the
asymptotic behaviour of solutions at infinity where the fluid density
becomes negligible. Here we recover vacuum solutions corresponding to
the slow and fast branches of O'Hanlon and Tupper's vacuum
solutions~\cite{OHT72} for $k=0$, and the Milne universe for $k<0$
models.  The expanding Milne universe is a late-time attractor (and
the collapsing solution is an early-time atractor) for $\gamma>2/3$.

We have seen that it is only possible to construct solutions with
infinite proper lifetimes for values of the Brans-Dicke parameter
$\omega<-4/3$. However, in all cases the solutions possess a curvature
singularity in the conformally related Eisntein frame.

\section*{Acknowledgements}

The authors are grateful to John Barrow, Roy Maartens and David
Matravers for helpful comments.
DJH acknowledges financial support from the Defence Evaluation Research 
Agency.

\end{document}